\documentstyle[11pt]{article}
\begin{document}
\title{Constructing Doubly Self-Dual Chiral p-Form Actions in D=2(p+1) 
 Spacetime Dimensions}
\author{{Yan-Gang Miao${}^{{\rm a,b},1}$, Harald J.W. M$\ddot{\rm u}$ller-Kirsten${}^{{\rm a},2}$ 
and Dae Kil Park${}^{{\rm c,d},3}$}\\
{\small ${}^{\rm a}$ Department of Physics, University of Kaiserslautern,
P.O. Box 3049,}\\
{\small D-67653 Kaiserslautern, Germany}\\
{\small ${}^{\rm b}$ Department of Physics, Xiamen University, Xiamen 361005,}\\{\small People's Republic of 
China}\\
{\small ${}^{\rm c}$ Department of Physics, Kyungnam University, Masan 631-701, Korea}\\
{\small ${}^{\rm d}$ Michigan Center for Theoretical Physics, Randall 
Laboratory,}\\
{\small Department of Physics, University of Michigan,}\\
{\small Ann Arbor, MI 
48109-1120, USA}}
\date{}
\maketitle
\footnotetext[1]{Email address: miao@physik.uni-kl.de}
\footnotetext[2]{Email address: mueller1@physik.uni-kl.de}
\footnotetext[3]{Email address: dkpark@hep.kyungnam.ac.kr}
\vskip 18pt
\begin{center}{\bf Abstract}\end{center}
\baselineskip 22pt
\par
A Siegel-type chiral {\em p}-form action is proposed in $D=2(p+1)$ spacetime
dimensions. The approach we adopt is to realize
the symmetric 
second-rank Lagrange-multiplier field, introduced in Siegel's action, 
in terms of 
a normalized multiplication of two ({\em q}+1)-form fields with 
{\em q} indices of each field contracted in the even {\em p} case, or
of two pairs of
({\em q}+1)-form fields with {\em q} indices of each pair of fields
contracted in the odd {\em p} case, where the ({\em q}+1)-form fields 
are of external derivatives of one auxiliary {\em q}-form field for the former,
or of a pair of auxiliary {\em q}-form fields for the latter. Using 
this action, it is straightforward to deduce the recently constructed
PST action for {\em q} equal to zero.
It is found
that the Siegel-type chiral {\em p}-form action with a fixed {\em p} 
(even or odd) is doubly self-dual in $D=2(p+1)$ spacetime dimensions when the 
auxiliary field(s) is/are also chosen to be of {\em p}-form. 
This result includes PST's as a special case where 
{\em only} the chiral 0-form action is doubly self-dual in $D=2$ dimensions.
\vskip 24pt
PACS number(s): 11.10.-z, 11.15.-q, 11.30.-j
\newpage
\section{Introduction}
\par
Chiral {\em p}-forms exist in the $D=2(p+1)$ dimensional spacetime since their
external derivatives, {\em i.e.}, the field strengths, are ({\em p}+1)-forms
and the Hodge duals as well, which requires the spacetime dimensions to be
twice big the form number of the field strengths. The chirality usually
means that the field strengths satisfy a self-duality condition. In 
spacetime with Lorentzian metric signature, self-duality requires the
chiral {\em p}-forms to be complex if {\em p} is odd. In this case one may
equivalently
introduce [1] a pair of real {\em p}-forms, $A^{a(p)}(x), a=1,2$, and
impose upon the corresponding field strengths, $F^{a(p+1)}(A)=dA^{a(p)}(x)$, 
the duality
condition\footnote[1]{If $A^{a(p)}(x)$ are treated as an O(2) doublet and an
extended dualization is defined, the extended duals of $F^{a(p+1)}(A)$, if we
choose
$-{\epsilon}^{ab}{}^{\ast}F^{b(p+1)}(A)$, satisfy a self-duality condition.
See, for example, Ref.[2] for details.},
$$
{\cal F}^{a(p+1)}(A) \equiv {\epsilon}^{ab}F^{b(p+1)}(A)
-{}^{\ast}F^{a(p+1)}(A)=0, \hspace{15pt}
{\cal F}^{a(p+1)}(A)={\epsilon}^{ab}{}^{\ast}{\cal F}^{b(p+1)}(A),
$$
where ${\epsilon}^{12}=-{\epsilon}^{21}=1, a,b=1,2$, and ${}^{\ast}F^{a(p+1)}
(A)$ stand for the Hodge duals of $F^{a(p+1)}(A)$.
However, the self-duality requires the chiral {\em p}-forms to be real if 
{\em p} is even. For this case one {\em p}-form potential, $A^{(p)}(x)$, is 
enough\footnote[2]{We do not introduce two {\em p}-form potentials that 
constitute a
doublet adopted in Ref.[2] because that treatment is not economical for the 
even
{\em p} case.}, and its field strength, $F^{(p+1)}(A)=dA^{(p)}(x)$, 
of course, satisfies 
the usual 
self-duality condition,
$$
{\cal F}^{(p+1)}(A) \equiv F^{(p+1)}(A)
-{}^{\ast}F^{(p+1)}(A)=0, \hspace{15pt}
{\cal F}^{(p+1)}(A)=-{}^{\ast}{\cal F}^{(p+1)}(A).
$$
For the sake of conciseness, we combine these two (self-)duality conditions
into one general form\footnote[3]{For the definition of ${\Gamma}^{ab}$, see
eq.(10).} for both even {\em p} and odd {\em p} cases,
\begin{equation}
{\cal F}^{a(p+1)}(A) \equiv {\Gamma}^{ab}F^{b(p+1)}(A)
-{}^{\ast}F^{a(p+1)}(A)=0, \hspace{15pt}
{\cal F}^{a(p+1)}(A)=(-1)^{p+1}{\Gamma}^{ab}{}^{\ast}{\cal F}^{b(p+1)}(A).
\end{equation}

For the main reason why chiral {\em p}-forms have received so much attention,
we may summarize that they appear in various theoretical models that relate to
superstring theories, and reflect especially the existence of a variety of
important dualities that connect these theories among one another.

One has to envisage the two basic problems in a Lagrangian description of 
chiral {\em p}-forms: one is the consistent quantization and the other is the
harmonic combination of manifest duality and spacetime covariance, since the
equation of motion of a chiral {\em p}-form, {\em i.e.}, the
self-duality condition, is first order with respect to the derivatives of
space and time. In order to solve these problems, some non-manifestly spacetime
covariant [1,3-8] and manifestly spacetime covariant [9-12] models have
been proposed. 
It is remarkable that these chiral {\em p}-form models have close 
relationships 
among one another, especially
various dualities that
have been
demonstrated in detail from the points of view of both configuration [12-15]
and
momentum [16,17] spaces. 

The recently constructed PST action [12], motivated by the SS action [8]
together with the idea [18] of spacetime covariance in it, has become a
successful model in putting both the manifest spacetime and duality symmetries
into a chiral {\em p}-form action. The characteristic of this model is to
introduce a Lagrangian-multiplier term in a non-polynomial way, or more
precisely,
in terms of a normalized multiplication of two vector (1-form) fields 
that
are ``external'' derivatives of one auxiliary scalar (0-form) field. 
Though there is the merit
of connecting the Zwanziger [1] with SS actions or with others through
various dualities, it is quite unsatisfactory that the model lacks of [13] 
double self-duality 
in higher (than two) dimensions. The double
self-duality means 
that the action of chiral {\em p}-forms is self-dual with respect 
{\em not only} to dualization of 
chiral fields, {\em but also} to dualization of auxiliary 
fields. We may ask whether the double self-duality does or does not appear in
higher (than two) dimensions.

Analysing in detail Siegel's action\footnote[4]{There is a recent paper [19]
where a so-called generalized gauged Siegel model is developed. This
model and its related discussions on interference phenomena and Lorentz
invariance are limited only in $D=2$ dimensions. However, our attention here
focuses on beyond $D=2$ dimensions.}
[9] and its generalization
[17] in $D=2(p+1)$ ({\em p} even) dimensions, we immediately find the relation
between the Siegel and the PST actions. That is, the former becomes the 
latter
if the symmetric second-rank Lagrangian-multiplier is realized in terms of the
approach mentioned above. In fact, this discovery relates closely to 
our question. As we have proven,
if the Lagrangian-multiplier is realized in terms of higher than zero
form auxiliary fields, the double self-duality can occur in higher
(than two) dimensions. In this way, we generalize PST's result deduced by the
 assumption of 
introducing one and the same auxiliary scalar field in any even dimensions.
 
 It is the aim of this paper to construct a doubly self-dual chiral 
{\em p}-form action 
in $D=2(p+1)$ spacetime dimensions. 
The paper is arranged as follows. In the next section a Siegel-type
chiral {\em p}-form action is proposed in $D=2(p+1)$ dimensions. The 
  method we utilize is to realize the symmetric second-rank 
Lagrange-multiplier field of Siegel's action
in terms of a normalized multiplication of two ({\em q}+1)-form fields with
{\em q} indices of each field contracted for the even {\em p} case\footnote[5]
{Zero is included in the even {\em p} case throughout this paper.}, or of two
pairs of ({\em q}+1)-form fields with {\em q} indices of each pair of fields
contracted for the odd {\em p} case, where the ({\em q}+1)-form fields
are of external derivatives of one auxiliary {\em q}-form field for the former,
or of one pair of auxiliary {\em q}-form fields for the latter. In section 3
we investigate duality properties of this action with respect to {\em both}
the dualization of chiral {\em p}-form fields {\em and} the dualization of
auxiliary {\em q}-form fields, and make search for the condition when
the
double self-duality appears.
Finally section 4 is 
devoted to a conclusion.

   The metric notation we use throughout this paper is
\begin{eqnarray}
& &g_{00}=-g_{11}=\cdots=-g_{D-1,D-1}=1,\nonumber\\ 
& &{\epsilon}^{012{\cdots}D-1}=1.
\end{eqnarray}
Greek letters stand for spacetime indices (${\mu},{\nu},{\sigma},{\rho},
{\cdots}=0,1,2,\cdots,
D-1$).
\section{Siegel-type chiral p-form action in D=2(p+1) dimensions}
We begin with Siegel's action in $D=2(p+1)$ ({\em p} even) dimensions [17],
\begin{eqnarray}
S_{s}=\int d^{D}x \left\{-\frac{1}{2(p+1)!}
{F}_{{\mu}_{1}
{\cdots}{\mu}_{p+1}}(A)
{F}^{{\mu}_{1}
{\cdots}{\mu}_{p+1}}(A)
+\frac{1}{2}{{\lambda}^{\mu}}_{\nu}
{\cal F}_{{\mu}{\mu}_{1}
{\cdots}{\mu}_{p}}(A)
{\cal F}^{{\nu}{\mu}_{1}
{\cdots}{\mu}_{p}}(A)\right\},
\end{eqnarray}
where 
$A_{{\mu}_{1}{\cdots}{\mu}_{p}}(x)$ 
is a real {\em p}-form field,
$F_{{\mu}_{1}{\cdots}{\mu}_{p+1}}(A)$ 
its field strength, and
${\cal F}_{{\mu}_{1}{\cdots}{\mu}_{p+1}}(A)$, which is called the self-dual
tensor in Refs.[12,13], is 
the difference of the field strength and its Hodge dual. Note that the
Lagrange-multiplier,
${\lambda}_{{\mu}{\nu}}(x)$, 
is in general a symmetric second-rank tensor field. We may introduce a real
auxiliary {\em q}-form field,
$Y_{{\mu}_{1}{\cdots}{\mu}_{q}}(x)$,
and realize
${\lambda}_{{\mu}{\nu}}(x)$ in some sense in terms of a normalized
multiplication of two ({\em q}+1)-form fields with {\em q} indices of each
field contracted, where the ({\em q}+1)-form field,
$T_{{\mu}_{1}{\cdots}{\mu}_{q+1}}(x)$,
is defined as the external derivative of the {\em q}-form field, {\em i.e.},
\begin{equation}
{\lambda}_{{\mu}{\nu}}=\frac{ 
{T_{\mu}}^{{\sigma}_{1}{\cdots}{\sigma}_{q}}
T_{{\nu}{\sigma}_{1}{\cdots}{\sigma}_{q}}}
 {T^2},
\end{equation}
where
\begin{eqnarray}
T_{{\mu}_{1}{\cdots}{\mu}_{q+1}} &\equiv&
{\partial}_{[{\mu}_{1}}Y_{{\mu}_{2}{\cdots}{\mu}_{q+1}]},\nonumber\\
T^2 &\equiv&
T_{{\mu}_{1}{\cdots}{\mu}_{q+1}}
T^{{\mu}_{1}{\cdots}{\mu}_{q+1}}. 
\end{eqnarray}
Here we mention that {\em q} takes numbers $0,1,\cdots,D-2=2p$ for a fixed
{\em p} (even or odd) because the dual of a {\em q}-form is a
$(D-q-2)$-form, {\em i.e.}, a
$(2p-q)$-form in $D=2(p+1)$ spacetime dimensions. For details, we refer to
the next section. Substituting eq.(4) into eq.(3), one obtains a 
Siegel-type action for chiral {\em p}-forms in $D=2(p+1)$ ({\em p} even) 
dimensions, 
\begin{eqnarray} 
S_{st}=\int \lefteqn{d^{D}x \left\{
-\frac{1}{2(p+1)!}
{F}_{{\mu}_{1}
{\cdots}{\mu}_{p+1}}(A)
{F}^{{\mu}_{1}
{\cdots}{\mu}_{p+1}}(A)\right.}\nonumber\\
& & \left.\hspace{5pt}
+\frac{ 
T^{{\mu}{\sigma}_{1}{\cdots}{\sigma}_{q}}
T_{{\nu}{\sigma}_{1}{\cdots}{\sigma}_{q}}}
 {2T^2}
{\cal F}_{{\mu}{\mu}_{1}
{\cdots}{\mu}_{p}}(A)
{\cal F}^{{\nu}{\mu}_{1}
{\cdots}{\mu}_{p}}(A)\right\}.
\end{eqnarray}

When {\em p} is odd, we may introduce a pair of real {\em p}-form
fields,
$A^{a}_{{\mu}_{1}{\cdots}{\mu}_{p}}(x), a=1,2$, 
as explained in section 1. Correspondingly, we also introduce a pair of real
auxiliary {\em q}-form fields,
$Y^{b}_{{\mu}_{1}{\cdots}{\mu}_{q}}(x), b=1,2.$
Therefore, in accordance with eq.(6), we conclude that the Siegel-type 
action that combines {\em both} even {\em p} 
{\em and} odd {\em p} cases takes the form,
\begin{eqnarray} 
S_{ST}=\int \lefteqn{d^{D}x \bigg\{
-\frac{1}{2{\Delta}(p+1)!}
{F}^{a}_{{\mu}_{1}
{\cdots}{\mu}_{p+1}}(A)
{F}^{a{\mu}_{1}
{\cdots}{\mu}_{p+1}}(A)}\nonumber\\
& & \hspace{5pt}+
\frac{ 
T^{b{\mu}{\sigma}_{1}{\cdots}{\sigma}_{q}}
T^{b}_{{\nu}{\sigma}_{1}{\cdots}{\sigma}_{q}}}
 {2{\Delta}T^2}
{\cal F}^{a}_{{\mu}{\mu}_{1}
{\cdots}{\mu}_{p}}(A)
{\cal F}^{a{\nu}{\mu}_{1}
{\cdots}{\mu}_{p}}(A)\bigg\},
\end{eqnarray}
where
\begin{eqnarray}
F^{a}_{{\mu}_{1}
{\cdots}{\mu}_{p+1}}(A) &\equiv&
{\partial}_{[{\mu}_{1}}
A^{a}_{{\mu}_{2}{\cdots}{\mu}_{p+1}]},\nonumber\\
{\cal F}^{a}_{{\mu}_{1}{\cdots}{\mu}_{p+1}}(A) &\equiv& {\Gamma}^{ab}
F^{b}_{{\mu}_{1}{\cdots}{\mu}_{p+1}}(A)
-\frac{1}{(p+1)!}{\epsilon}_{{\mu}_{1}{\cdots}{\mu}_{p+1}{\nu}_{1}{\cdots}
{\nu}
_{p+1}}
F^{a{\nu}_{1}
{\cdots}{\nu}_{p+1}}(A),\\
T^{a}_{{\mu}_{1}{\cdots}{\mu}_{q+1}} &\equiv&
{\partial}_{[{\mu}_{1}}Y^{a}_{{\mu}_{2}{\cdots}{\mu}_{q+1}]},\nonumber\\
T^2 &\equiv&
T^{a}_{{\mu}_{1}{\cdots}{\mu}_{q+1}}
T^{a{\mu}_{1}{\cdots}{\mu}_{q+1}},
\end{eqnarray}
and
\begin{eqnarray}
{\Delta}&=&\left\{\begin{array}{lr}
1, & {\rm for\hspace{5pt}even} \hspace{5pt}p, \\
2, & {\rm for\hspace{5pt}odd} \hspace{5pt}p, 
\end{array}\right.\nonumber\\
{\Gamma}^{ab}&=&\left\{\begin{array}{lcr}
{\delta}^{ab}=1, & a=b=1, & {\rm for\hspace{5pt}even} \hspace{5pt}p, \\
{\epsilon}^{ab}, & a,b=1,2, \hspace{3pt}{\rm and}\hspace{5pt} 
{\epsilon}^{12}=-{\epsilon}^{21}=1, &
{\rm for\hspace{5pt}odd} \hspace{5pt}p.
\end{array}\right.
\end{eqnarray}
In addition, one obtains from eq.(8) that for the both cases the field 
strength differences, or the so-called self-dual tensors, satisfy the
relations
\begin{equation}
{\cal F}^{a}_{{\mu}_{1}{\cdots}{\mu}_{p+1}}(A)=
\frac{(-1)^{p+1}}{(p+1)!}
{\Gamma}^{ab}
{\epsilon}_{{\mu}_{1}{\cdots}{\mu}_{p+1}{\nu}_{1}{\cdots}
{\nu}
_{p+1}}
{\cal F}^{b{\nu}_{1}
{\cdots}{\nu}_{p+1}}(A),
\end{equation}
which will be useful in calculations in the next section. These relations 
are the
generalization of the special forms discussed in $D=2,4,6$ 
dimensions [12,13].

We point out that in the even {\em p} case the Siegel-type action, eq.(7) or
eq.(6), contains
the PST action in $D=2$ and $D=6$ dimensions 
when {\em q} equals zero, and in the odd {\em p} case eq.(7) includes 
the PST
action in $D=4$ dimensions if only one auxiliary 0-form field is introduced. 
Consequently, 
the PST action can
be treated as a special example of the Siegel-type action. Furthermore, due to
the fact that the PST action originates from the SS action
[8] which is dual to the Zwanziger action [1], our discovery establishes 
the relationship between the Siegel-type and the SS or Zwanziger
actions through the PST action as a bridge.
\section{Duality properties of the chiral p-form action in D=2(p+1)
dimensions}
In the following we investigate duality properties of the Siegel-type chiral
{\em p}-form action, $S_{ST}$, with respect to {\em both} the dualization 
of the
chiral {\em p}-form fields, 
$A^{a}_{{\mu}_{1}{\cdots}{\mu}_{p}}(x)$, {\em and} the dualization 
of the auxiliary
{\em q}-form fields,
$Y^{b}_{{\mu}_{1}{\cdots}{\mu}_{q}}(x)$,
and make search for the condition under which double self-duality occurs. 
For the dualization of the chiral fields we 
shall
follow the usual line of Refs.[13-15], 
but
for the dualization of the auxiliary fields we shall proceed in
our own way with the intention
to avoid extensive calculations and to arrive directly at the desired
result.
\subsection{Dualization of the chiral p-form fields}
By introducing two ({\em p} even) or two pairs of ({\em p} odd)
independent ({\em p}+1)-form fields, $ 
F^{a}_{{\mu}_{1}
{\cdots}{\mu}_{p+1}}$
and $
G^{b}_{{\mu}_{1}
{\cdots}{\mu}_{p+1}}$,
we construct a new
action to replace $S_{ST}$,
\begin{eqnarray} 
S_{ST}^{\prime}&=&\int d^{D}x \bigg\{
-\frac{1}{2{\Delta}(p+1)!}
{F}^{a}_{{\mu}_{1}
{\cdots}{\mu}_{p+1}}
{F}^{a{\mu}_{1}
{\cdots}{\mu}_{p+1}}
+\frac{ 
T^{b{\mu}{\sigma}_{1}{\cdots}{\sigma}_{q}}
T^{b}_{{\nu}{\sigma}_{1}{\cdots}{\sigma}_{q}}}
 {2{\Delta}T^2}
{\cal F}^{a}_{{\mu}{\mu}_{1}
{\cdots}{\mu}_{p}}
{\cal F}^{a{\nu}{\mu}_{1}
{\cdots}{\mu}_{p}}\nonumber\\
& & \hspace{35pt}+\frac{1}{{\Delta}(p+1)!}
G^{a{\mu}_{1}
{\cdots}{\mu}_{p+1}}\left(
F^{a}_{{\mu}_{1}
{\cdots}{\mu}_{p+1}}-
{\partial}_{[{\mu}_{1}}
A^{a}_{{\mu}_{2}{\cdots}{\mu}_{p+1}]}\right)\bigg\},
\end{eqnarray}
where 
$
{\cal F}^{a{\mu}_{1}{\cdots}{\mu}_{p+1}}
$
has the same definition as that of
$
{\cal F}^{a{\mu}_{1}{\cdots}{\mu}_{p+1}}(A)
$
in eq.(8), but here it is not dealt with as a functional of
$
A^{a}_{{\mu}_{1}{\cdots}{\mu}_{p}}(x)
$.
Variation of
$S_{ST}^{\prime}$ with respect to
$ 
G^{a{\mu}_{1}
{\cdots}{\mu}_{p+1}}
$
gives
\begin{equation}
F^{a}_{{\mu}_{1}
{\cdots}{\mu}_{p+1}}=
{\partial}_{[{\mu}_{1}}
A^{a}_{{\mu}_{2}{\cdots}{\mu}_{p+1}]},
\end{equation}
which, when substituted into $S_{ST}^{\prime}$, yields the classical 
equivalence between the two actions, $S_{ST}$ and
$S_{ST}^{\prime}$. Furthermore, variation of $S_{ST}^{\prime}$ with respect to
$ 
F^{a}_{{\mu}_{1}
{\cdots}{\mu}_{p+1}}
$
leads to the expression of
$
G^{a{\mu}_{1}
{\cdots}{\mu}_{p+1}}
$
in terms of 
$
F^{a{\mu}_{1}
{\cdots}{\mu}_{p+1}}
$ and other fields,
\begin{eqnarray}
G^{a{\mu}_{1}
{\cdots}{\mu}_{p+1}}&=&
F^{a{\mu}_{1}
{\cdots}{\mu}_{p+1}}
-\frac{1}{T^2}\left\{\phantom{\frac{1}{1}}{\Gamma}^{ab}T^{c{\sigma}_{1}{\cdots}
{\sigma}_{q}[{\mu}_{1}}
{\cal F}^{{\mu}_{2}{\cdots}{\mu}_{p+1}]{\sigma}_{q+1}b}T^{c}_{{\sigma}_{1}
{\cdots}
{\sigma}_{q+1}}\right.\nonumber\\
& & + \left. \frac{1}{(p+1)!}{\epsilon}^{{\mu}_{1}{\cdots}
{\mu}_{p+1}{\nu}_{1}
{\cdots}
{\nu}_{p+1}}T^{b}_{{\sigma}_{1}{\cdots}{\sigma}_{q}[{\nu}_{1}}
{\cal F}^{a}_{{\nu}
_{2}{\cdots}{\nu}_{p+1}]{\sigma}_{q+1}}T^{b{\sigma}_{1}{\cdots}{\sigma}_{q+1}}
\right\}.
\end{eqnarray}
If one 
defines another ({\em p} even) or another pair of ({\em p} odd) field 
strength difference(s) that is/are relevant to
$
G^{a{\mu}_{1}
{\cdots}{\mu}_{p+1}}
$,
\begin{equation}
{\cal G}^{a{\mu}_{1}{\cdots}{\mu}_{p+1}} \equiv {\Gamma}^{ab}
G^{b{\mu}_{1}{\cdots}{\mu}_{p+1}}
-\frac{1}{(p+1)!}{\epsilon}^{{\mu}_{1}{\cdots}{\mu}_{p+1}{\nu}_{1}{\cdots}
{\nu}
_{p+1}}
G^{a}_{{\nu}_{1}
{\cdots}{\nu}_{p+1}},
\end{equation}
and substitutes eq.(14) into eq.(15), one then establishes the relation(s)
between the two ({\em p} even) or two pairs of ({\em p} odd) field 
strength differences,
\begin{equation}
{\cal F}^{a{\mu}_{1}{\cdots}{\mu}_{p+1}}=
{\cal G}^{a{\mu}_{1}{\cdots}{\mu}_{p+1}}.
\end{equation}
 One may note that such a relationship exists in
various chiral {\em p}-form actions, which has been pointed out 
in {\em both} configuration [14] {\em and} momentum [17] 
spaces. 
With
the aid of eq.(16), one can easily obtain from eq.(14) 
$
F^{a{\mu}_{1}
{\cdots}{\mu}_{p+1}}
$
expressed in
terms of 
$
G^{a{\mu}_{1}
{\cdots}{\mu}_{p+1}}
$ and other fields,
\begin{eqnarray}
F^{a{\mu}_{1}
{\cdots}{\mu}_{p+1}}&=&
G^{a{\mu}_{1}
{\cdots}{\mu}_{p+1}}
+\frac{1}{T^2}\left\{\phantom{\frac{1}{1}}{\Gamma}^{ab}T^{c{\sigma}_{1}{\cdots}
{\sigma}_{q}[{\mu}_{1}}
{\cal G}^{{\mu}_{2}{\cdots}{\mu}_{p+1}]{\sigma}_{q+1}b}T^{c}_{{\sigma}_{1}
{\cdots}
{\sigma}_{q+1}}\right.\nonumber\\
& & + \left. \frac{1}{(p+1)!}{\epsilon}^{{\mu}_{1}{\cdots}{\mu}_{p+1}{\nu}_{1}
{\cdots}
{\nu}_{p+1}}T^{b}_{{\sigma}_{1}{\cdots}{\sigma}_{q}[{\nu}_{1}}
{\cal G}^{a}_{{\nu}
_{2}{\cdots}{\nu}_{p+1}]{\sigma}_{q+1}}T^{b{\sigma}_{1}{\cdots}{\sigma}_{q+1}}
\right\}.
\end{eqnarray}
We can check from eq.(14) that when the self-duality condition is satisfied, 
{\em i.e.}, 
$
{\cal F}^{a{\mu}_{1}{\cdots}{\mu}_{p+1}}=0,
$
which is also called an ``on mass shell'' condition,
$
F^{a{\mu}_{1}
{\cdots}{\mu}_{p+1}}
$
and
$
G^{b{\mu}_{1}
{\cdots}{\mu}_{p+1}}
$
relate with a duality, 
\begin{equation}
F^{a{\mu}_{1}
{\cdots}{\mu}_{p+1}}=\frac{(-1)^p}{(p+1)!}{\Gamma}^{ab}{\epsilon}^{{\mu}_{1}
{\cdots}{\mu}_{p+1}{\nu}_{1}{\cdots}{\nu}_{p+1}}G^{b}_{{\nu}_{1}{\cdots}
{\nu}_{p+1}}.
\end{equation}
Substituting eq.(17) into eq.(12) and dropping a total derivative term, 
we obtain the dual action of $S_{ST}$,
\begin{eqnarray} 
S_{ST}^{dual}&=& \int d^{D}x \bigg\{
\frac{1}{2{\Delta}(p+1)!}
{G}^{a}_{{\mu}_{1}
{\cdots}{\mu}_{p+1}}
{G}^{a{\mu}_{1}
{\cdots}{\mu}_{p+1}}
+\frac{ 
T^{b{\mu}{\sigma}_{1}{\cdots}{\sigma}_{q}}
T^{b}_{{\nu}{\sigma}_{1}{\cdots}{\sigma}_{q}}}
 {2{\Delta}T^2}
{\cal G}^{a}_{{\mu}{\mu}_{1}
{\cdots}{\mu}_{p}}
{\cal G}^{a{\nu}{\mu}_{1}
{\cdots}{\mu}_{p}}\nonumber\\
& & \hspace{35pt}+\frac{1}{{\Delta}}
A^{a}_{{\mu}_{1}{\cdots}{\mu}_{p}}
{\partial}_{{\mu}}
G^{a{\mu}{\mu}_{1}
{\cdots}{\mu}_{p}}
\bigg\}.
\end{eqnarray}
Variation of eq.(19) with respect to 
$
A^{a}_{{\mu}_{1}{\cdots}{\mu}_{p}}(x)
$
gives
$
{\partial}_{{\mu}}
G^{a{\mu}{\mu}_{1}
{\cdots}{\mu}_{p}}=0,
$
whose solution has to be
\begin{eqnarray}
G^{a{\mu}_{1}
{\cdots}{\mu}_{p+1}}&=&
\frac{(-1)^p}{(p+1)!}{\Gamma}^{ab}{\epsilon}^{{\mu}_{1}
{\cdots}{\mu}_{p+1}{\nu}_{1}{\cdots}{\nu}_{p+1}}
{\partial}_{[{\nu}_{1}}
B^{b}_{{\nu}_{2}{\cdots}{\nu}_{p+1}]}\nonumber\\
&\equiv&
\frac{(-1)^p}{(p+1)!}{\Gamma}^{ab}{\epsilon}^{{\mu}_{1}
{\cdots}{\mu}_{p+1}{\nu}_{1}{\cdots}{\nu}_{p+1}}
F^{b}_{{\nu}_{1}{\cdots}
{\nu}_{p+1}}(B),
\end{eqnarray}
where 
$
B^{a}_{{\mu}_{1}{\cdots}{\mu}_{p}}(x)
$ is one ({\em p} even) or are a pair of ({\em p} odd)
arbitrary {\em p}-form field(s) we introduce.
Substituting eq.(20) into eq.(19), one finally obtains 
the dual action in terms
of 
$
B^{a}_{{\mu}_{1}{\cdots}{\mu}_{p}}(x)
$ and other fields,
\begin{eqnarray} 
S_{ST}^{dual}=\int \lefteqn{d^{D}x \bigg\{
-\frac{1}{2{\Delta}(p+1)!}
{F}^{a}_{{\mu}_{1}
{\cdots}{\mu}_{p+1}}(B)
{F}^{a{\mu}_{1}
{\cdots}{\mu}_{p+1}}(B)}\nonumber\\
& & \hspace{5pt}+
\frac{ 
T^{b{\mu}{\sigma}_{1}{\cdots}{\sigma}_{q}}
T^{b}_{{\nu}{\sigma}_{1}{\cdots}{\sigma}_{q}}}
 {2{\Delta}T^2}
{\cal F}^{a}_{{\mu}{\mu}_{1}
{\cdots}{\mu}_{p}}(B)
{\cal F}^{a{\nu}{\mu}_{1}
{\cdots}{\mu}_{p}}(B)\bigg\}.
\end{eqnarray}

This action, eq.(21), has the same form as the original one, eq.(7), only with 
the
replacement of 
$
A^{a}_{{\mu}_{1}{\cdots}{\mu}_{p}}(x)
$
by 
$
B^{a}_{{\mu}_{1}{\cdots}{\mu}_{p}}(x)
$.
As analysed above, 
$
A^{a}_{{\mu}_{1}{\cdots}{\mu}_{p}}(x)
$
and
$
B^{b}_{{\mu}_{1}{\cdots}{\mu}_{p}}(x)
$
coincide with each other up to a constant when the self-duality
condition is imposed. Consequently, the Siegel-type chiral {\em p}-form
action is self-dual with respect to 
$
A^{a}_{{\mu}_{1}{\cdots}{\mu}_{p}}(x)
-
B^{b}_{{\mu}_{1}{\cdots}{\mu}_{p}}(x)
$
dualization given
by eqs.(13), (18) and (20).
\subsection{Dualization of the auxiliary q-form fields}
For the sake of convenience in the following discussion, we rewrite eq.(7) to
be
\begin{eqnarray} 
S_{ST}&=&\int d^{D}x \lefteqn{\bigg\{
-\frac{1}{2{\Delta}(p+1)!}
{F}^{a}_{{\mu}_{1}
{\cdots}{\mu}_{p+1}}(A)
{F}^{a{\mu}_{1}
{\cdots}{\mu}_{p+1}}(A)}\nonumber\\
& & \hspace{35pt}+
\frac{1}{2{\Delta}}
\frac{
{\partial}^{[{\mu}}Y^{{\sigma}_{1}{\cdots}{\sigma}_{q}]b}
{\partial}_{[{\nu}}Y^{b}_{{\sigma}_{1}{\cdots}{\sigma}_{q}]}
} 
{
{\partial}_{[{\nu}_{1}}Y^{c}_{{\nu}_{2}{\cdots}{\nu}_{q+1}]}
{\partial}^{[{\nu}_{1}}Y^{{\nu}_{2}{\cdots}{\nu}_{q+1}]c}}
{\cal F}^{a}_{{\mu}{\mu}_{1}
{\cdots}{\mu}_{p}}(A)
{\cal F}^{a{\nu}{\mu}_{1}
{\cdots}{\mu}_{p}}(A)
\bigg\}.
\end{eqnarray}
\par
We introduce two (for even {\em p}) or two pairs of (for odd {\em p}) 
({\em q}+1)-form fields,
$
U^{a}_{{\mu}_{1}{\cdots}{\mu}_{q+1}}(x)
$
and
$
V^{b}_{{\mu}_{1}{\cdots}{\mu}_{q+1}}(x)
$,
and replace eq.(22) by the action,
\begin{eqnarray} 
S_{ST}^{\prime}&=&\int d^{D}x \lefteqn{\left\{
-\frac{1}{2{\Delta}(p+1)!}
{F}^{a}_{{\mu}_{1}
{\cdots}{\mu}_{p+1}}(A)
{F}^{a{\mu}_{1}
{\cdots}{\mu}_{p+1}}(A)\right.}\nonumber\\
& & \hspace{35pt}+
\frac{1}{2{\Delta}}
\frac{
{U}^{b{\mu}{\sigma}_{1}{\cdots}{\sigma}_{q}}
{U}^{b}_{{\nu}{\sigma}_{1}{\cdots}{\sigma}_{q}}
} 
{
{U}^{c}_{{\nu}_{1}{\cdots}{\nu}_{q+1}}
{U}^{c{\nu}_{1}{\cdots}{\nu}_{q+1}}}
{\cal F}^{a}_{{\mu}{\mu}_{1}
{\cdots}{\mu}_{p}}(A)
{\cal F}^{a{\nu}{\mu}_{1}
{\cdots}{\mu}_{p}}(A)\nonumber\\
& & \hspace{35pt}\left.+
\frac{1}{{\Delta}(q+1)!}
V^{a{\mu}_{1}{\cdots}{\mu}_{q+1}}\left(
U^{a}_{{\mu}_{1}{\cdots}{\mu}_{q+1}}-
{\partial}_{[{\mu}_{1}}Y^{a}_{{\mu}_{2}{\cdots}{\mu}_{q+1}]}
\right)\right\},
\end{eqnarray}
where
$
U^{a}_{{\mu}_{1}{\cdots}{\mu}_{q+1}}(x)
$
and
$
V^{b}_{{\mu}_{1}{\cdots}{\mu}_{q+1}}(x)
$
act,
at present, as independent auxiliary fields.
 Variation of eq.(23) with respect to 
$
V^{a{\mu}_{1}{\cdots}{\mu}_{q+1}}(x)
$
gives
\begin{equation}
U^{a}_{{\mu}_{1}{\cdots}{\mu}_{q+1}}=
{\partial}_{[{\mu}_{1}}Y^{a}_{{\mu}_{2}{\cdots}{\mu}_{q+1}]},
\end{equation}
which yields the equivalence between the actions, eq.(22) and eq.(23). On the
other hand, variation of eq.(23) with respect to
$
U^{a}_{{\mu}_{1}{\cdots}{\mu}_{q+1}}(x)
$
leads to the expression of
$
V^{a{\mu}_{1}{\cdots}{\mu}_{q+1}}(x)
$
in terms of
$
U^{b{\mu}_{1}{\cdots}{\mu}_{q+1}}(x)
$
and $
{\cal F}^{c{\mu}_{1}
{\cdots}{\mu}_{p+1}}(A)$,
\begin{eqnarray}
V^{a{\mu}_{1}{\cdots}{\mu}_{q+1}}&=&(q+1)! \hspace{2pt}
U^{a{\mu}_{1}{\cdots}{\mu}_{q+1}}
\frac{
{U}^{b{\mu}{\sigma}_{1}{\cdots}{\sigma}_{q}}
{U}^{b}_{{\nu}{\sigma}_{1}{\cdots}{\sigma}_{q}}
} 
{\left(
{U}^{d}_{{\nu}_{1}{\cdots}{\nu}_{q+1}}
{U}^{d{\nu}_{1}{\cdots}{\nu}_{q+1}}\right)^2}
{\cal F}^{c}_{{\mu}{\rho}_{1}
{\cdots}{\rho}_{p}}(A)
{\cal F}^{c{\nu}{\rho}_{1}
{\cdots}{\rho}_{p}}(A)\nonumber\\
& & -
\frac{1}
{\left(
{U}^{c}_{{\nu}_{1}{\cdots}{\nu}_{q+1}}
{U}^{c{\nu}_{1}{\cdots}{\nu}_{q+1}}\right)}
{{\cal F}^{b}_{{\rho}_{1}{\cdots}{\rho}_{p}}}^{[{\mu}_{1}}(A)
{U_{{\rho}_{p+1}}}^{
{\mu}_{2}{\cdots}{\mu}_{q+1}]a}{\cal F}^{b{\rho}_{1}{\cdots}{\rho}_{p+1}}(A).
\end{eqnarray}
 Multiplying both sides of eq.(25) by
$
U^{a}_{{\mu}_{1}{\cdots}{\mu}_{q+1}}(x),
$
one obtains
\begin{equation}
U^{a}_{{\mu}_{1}{\cdots}{\mu}_{q+1}}
V^{a{\mu}_{1}{\cdots}{\mu}_{q+1}}=0.
\end{equation}
 This orthogonality relation between 
$
U^{a}_{{\mu}_{1}{\cdots}{\mu}_{q+1}}(x)
$
and
$
V^{a{\mu}_{1}{\cdots}{\mu}_{q+1}}(x)
$ generalizes the case of two vectors (1-forms) which are ``external''
derivatives of one and the same auxiliary scalar (0-form) adopted in Ref.[13].
Thus, the third term of eq.(23) reduces to
\begin{equation}
\frac{1}{\Delta}\int d^{D}x 
Y^{a}_{{\mu}_{1}{\cdots}{\mu}_{q}}{\partial}_{\mu}
V^{a{\mu}{\mu}_{1}{\cdots}{\mu}_{q}},
\end{equation}
where a total derivative contribution has been dropped.
Note that 
in the dual action of eq.(23) that is expressed in terms of 
$
F^{a}_{{\mu}_{1}{\cdots}{\mu}_{p+1}}(A)
$,
$
Y^{b}_{{\mu}_{1}{\cdots}{\mu}_{q}}(x)
$ and
$
V^{c}_{{\mu}_{1}{\cdots}{\mu}_{q+1}}(x)
$,
the first term of eq.(23) remains unchanged 
and the second retains no relation with the auxiliary {\em q}-form fields,
$
Y^{a}_{{\mu}_{1}{\cdots}{\mu}_{q}}(x)
$, because no such fields appear in eq.(25). Therefore, the variation of
the dual of eq.(23) with respect to
$
Y^{a}_{{\mu}_{1}{\cdots}{\mu}_{q}}(x)
$
is the same as the variation of eq.(23) itself with respect to
$
Y^{a}_{{\mu}_{1}{\cdots}{\mu}_{q}}(x)
$. This gives
$
{\partial}_{{\mu}}
V^{a{\mu}{\mu}_{1}{\cdots}{\mu}_{q}}=0
$,
whose solution has to be
\begin{equation}
V^{a{\mu}_{1}{\cdots}{\mu}_{q+1}}=
\frac{(-1)^p}{(2p-q+1)!}{\Gamma}^{ab}{\epsilon}^{{\mu}_{1}{\cdots}{\mu}_{q+1}
{\nu}_{1}{\cdots}
{\nu}_{2p-q+1}}
{\partial}_{[{\nu}_{1}}Z^{b}_{{\nu}_{2}{\cdots}{\nu}_{2p-q+1}]},
\end{equation}
where 
$
Z^{a}_{{\mu}_{1}{\cdots}{\mu}_{2p-q}}(x)
$
 is one ({\em p} even) or are a pair of ({\em p} odd) auxiliary $(2p-q)$-form 
field(s) that is/are the dual(s) of the
{\em q}-form(s),
$
Y^{b}_{{\mu}_{1}{\cdots}{\mu}_{q}}(x)
$,
in $D=2(p+1)$ dimensions.

In order for the dual of eq.(23) to possess self-duality, or putting it 
differently,
in order for eq.(22) to be
self-dual with respect to dualization of the auxiliary fields,
$
Y^{a}_{{\mu}_{1}{\cdots}{\mu}_{q}}(x)
$
and
$
Z^{b}_{{\mu}_{1}{\cdots}{\mu}_{2p-q}}(x)
$
must have the same form number, $q=2p-q$, {\em i.e.},
\begin{equation}
q=p.
\end{equation}
That is, the auxiliary fields have to be {\em p}-form. This is the condition
that the double self-duality happens.

 With eq.(29) one obtains from eq.(26) the dual relation between
$
V^{a{\mu}_{1}{\cdots}{\mu}_{p+1}}(x)
$
and
$
U^{b}_{{\mu}_{1}{\cdots}{\mu}_{p+1}}(x)
$,
\begin{equation}
V^{a{\mu}_{1}{\cdots}{\mu}_{p+1}} \propto {\Gamma}^{ab}
{\epsilon}^{{\mu}_{1}{\cdots}{\mu}_{p+1}{\nu}_{1}{\cdots}{\nu}_{p+1}}
U^{b}_{{\nu}_{1}{\cdots}{\nu}_{p+1}},
\end{equation}
where the proportionality coefficient, which is not explicitly shown, 
is in general a functional of
$
U^{a}_{{\mu}_{1}{\cdots}{\mu}_{p+1}}(x)
$ and
$
{\cal F}^{b}_{{\mu}_{1}{\cdots}{\mu}_{p+1}}(A)
$,
but with no indices left free. 
By using eq.(30),
one can prove the crucial relation,

{\small
\begin{equation}
\frac{
{U}^{b{\mu}{\sigma}_{1}{\cdots}{\sigma}_{p}}
{U}^{b}_{{\nu}{\sigma}_{1}{\cdots}{\sigma}_{p}}
} 
{
{U}^{c}_{{\nu}_{1}{\cdots}{\nu}_{p+1}}
{U}^{c{\nu}_{1}{\cdots}{\nu}_{p+1}}}
{\cal F}^{a}_{{\mu}{\mu}_{1}
{\cdots}{\mu}_{p}}(A)
{\cal F}^{a{\nu}{\mu}_{1}
{\cdots}{\mu}_{p}}(A)=-
\frac{
{V}^{b{\mu}{\sigma}_{1}{\cdots}{\sigma}_{p}}
{V}^{b}_{{\nu}{\sigma}_{1}{\cdots}{\sigma}_{p}}
} 
{
{V}^{c}_{{\nu}_{1}{\cdots}{\nu}_{p+1}}
{V}^{c{\nu}_{1}{\cdots}{\nu}_{p+1}}}
{\cal F}^{a}_{{\mu}{\mu}_{1}
{\cdots}{\mu}_{p}}(A)
{\cal F}^{a{\nu}{\mu}_{1}
{\cdots}{\mu}_{p}}(A),
\end{equation}}
where the identity,
\begin{equation}
{\cal F}^{a}_{{\mu}_{1}
{\cdots}{\mu}_{p+1}}(A)
{\cal F}^{a{\mu}_{1}
{\cdots}{\mu}_{p+1}}(A)=0,
\end{equation}
which can be deduced from eq.(11), was used. We note that it is
a hard job to determine the coefficient in eq.(30) even for the special cases
of $D=2,4,6$ dimensions as can be seen in Ref.[13]. However, it is interesting
that
we are not
obliged to determine the concrete form of the coefficient by following our
line of analyses shown from eq.(25) to eq.(32). 

Substituting eqs.(26), (29), (31) and the special case of eq.(28) with $q=p$ 
into eq.(23) and using eq.(32)
again,
one finally obtains the 
dual action in terms of the dual fields,
$
Z^{b}_{{\mu}_{1}{\cdots}{\mu}_{p}}(x),
$ and $
{F}^{a}_{{\mu}_{1}
{\cdots}{\mu}_{p+1}}(A)$,
\begin{eqnarray} 
S_{ST}^{dual}&=&\int d^{D}x \lefteqn{\bigg\{
-\frac{1}{2{\Delta}(p+1)!}
{F}^{a}_{{\mu}_{1}
{\cdots}{\mu}_{p+1}}(A)
{F}^{a{\mu}_{1}
{\cdots}{\mu}_{p+1}}(A)}\nonumber\\
& & \hspace{35pt}+
\frac{1}{2{\Delta}}
\frac{
{\partial}^{[{\mu}}Z^{{\sigma}_{1}{\cdots}{\sigma}_{p}]b}
{\partial}_{[{\nu}}Z^{b}_{{\sigma}_{1}{\cdots}{\sigma}_{p}]}
} 
{
{\partial}_{[{\nu}_{1}}Z^{c}_{{\nu}_{2}{\cdots}{\nu}_{p+1}]}
{\partial}^{[{\nu}_{1}}Z^{{\nu}_{2}{\cdots}{\nu}_{p+1}]c}}
{\cal F}^{a}_{{\mu}{\mu}_{1}
{\cdots}{\mu}_{p}}(A)
{\cal F}^{a{\nu}{\mu}_{1}
{\cdots}{\mu}_{p}}(A)
\bigg\}.
\end{eqnarray}
 As expected, the Siegel-type action with 
auxiliary {\em p}-form fields, $Y^{a}_{{\mu}_{1}{\cdots}{\mu}_{p}}(x)$, 
possesses 
self-duality 
with respect to dualization of the auxiliary
fields given by eqs.(24), (28), (29) and (30). Note that the dualization of the
auxiliary fields is ``off mass shell,'' which is different from the
dualization of the chiral {\em p}-form fields. This property is obviously
shown in eq.(25). We mention that the Siegel-type action is 
not self-dual with respect to dualization of auxiliary fields if the
condition, eq.(29), is not satisfied.
\section{Conclusion}
We have constructed the Siegel-type chiral {\em p}-form action by realizing
the symmetric 
second-rank Lagrange-multiplier field 
in terms of 
a normalized multiplication of two ({\em q}+1)-form fields with 
{\em q} indices of each field contracted in the even {\em p} case, or
of two pairs of
({\em q}+1)-form fields with {\em q} indices of each pair of fields
contracted in the odd {\em p} case, where the ({\em q}+1)-form fields 
are of external derivatives of one auxiliary {\em q}-form field of the former,
or of a pair of auxiliary {\em q}-form fields of the latter. 
 From this action, one can deduce the 
PST action in $D=2$ ($p=0$) and $D=6$ ($p=2$) dimensions simply by
letting {\em q} 
equal to zero, or the PST action in $D=4$ ($p=1$)
dimensions by
introducing only one auxiliary 0-form field. 
This means 
that the PST action can be treated as a special example of the Siegel-type
action. It is known [12-15] that the PST action originates from the SS 
action that is
dual to the Zwanziger action and that it has
close relations with others.  
For instance, on the one hand, it 
reduces to the
non-manifestly covariant FJ action [5,6] provided
appropriate gauge fixing conditions are chosen, on the other hand, it 
 becomes the
MWY action [11] if one gets rid of the PST action's non-polynomiality and
eliminates its scalar auxiliary field at the price of introducing 
{\em polynomially} 
auxiliary
({\em p}+1)-forms, 
or, vice versa, if one consistently 
truncates the 
MWY action's
infinite tail and puts at its end the auxiliary scalar field. Consequently, our
result shows that 
Siegel's action is in some sense the source of chiral {\em p}-form actions.

We have shown that the Siegel-type chiral {\em p}-form action is 
self-dual with respect to dualization of chiral {\em p}-form fields when
{\em q} assumes any of the numbers $0,1,\cdots,2p$, 
but it is
self-dual with respect to dualization of auxiliary {\em q}-form
fields {\em only} when {\em q} equals {\em p}, 
for a fixed {\em p} in $D=2(p+1)$ spacetime dimensions. 
As a consequence, 
$q=p$ is the condition under which the double self-duality appears.
This result includes PST's as a special case where 
{\em only} the chiral 0-form action is doubly self-dual in $D=2$ dimensions.

Finally, comment on the odd {\em p} case.

(i) The Siegel-type action is still self-dual with respect to dualization of 
chiral {\em p}-form fields even if only one auxiliary {\em q}-form field is 
introduced. In fact, this self-duality exists without any relation to concrete
realization of the Lagrange-multiplier field.

(ii) Regarding the self-duality with respect to dualization of auxiliary 
fields, 
we emphasize that it is
not enough to merely introduce one auxiliary {\em p}-form field. The reason
is that no
duality relation between 
$V_{{\mu}_{1}{\cdots}{\mu}_{p+1}}(x)$
and
$U^{{\mu}_{1}{\cdots}{\mu}_{p+1}}(x)$, like eq.(30), could be derived 
from eq.(26) with $q=p$ if there were only one auxiliary {\em p}-form field.
\vskip 33pt
\noindent
{\bf Acknowledgments}
\par
Y.-G. Miao 
acknowledges 
supports by an Alexander von Humboldt fellowship, by
the National Natural Science Foundation
of China under grant No.19705007, and by the Ministry of Education of China
under the special project for scholars returned from abroad. 
He 
would like to thank
D. Sorokin for correspondence. 
D.K. Park
acknowledges support from the Basic Research Program of the Korea Science 
and Engineering Foundation (Grant No. 2001-1-11200-001-2).

\end{document}